\documentclass[12pt,english]{article}
\usepackage[T1]{fontenc}
\usepackage[latin9]{inputenc}
\usepackage[a4paper]{geometry}
\usepackage{graphicx}



\usepackage{babel}

\begin{document}

\title{\textbf{The Rotation of Bulges}}
\date{}

\author{Dimitri A. Gadotti
\\
\\
European Southern Observatory, Casilla 19001, Santiago 19, Chile}
\maketitle

\begin{abstract}
I briefly review early and recent progress concerning the dynamical properties of bulges. I also show results from a study on the kinematics of bulges and disks in edge-on galaxies, where the bulge rotation curve is obtained with little contamination from the disk. I show that a proper structural analysis is crucial to understand such measurements. The results suggest that exponential bulges have a higher degree of rotational support than bulges with higher S\'ersic indexes, giving support to the concept that photometric bulges include different categories of stellar systems.
\end{abstract}

\section{Introduction}

Similarities between some (classical) bulges and massive elliptical galaxies lead to the idea that bulges are just scaled down ellipticals, and that both can form through violent processes. Although this is spread in the astronomical community, suggestions that such notion can in fact be (at least partially) a misconception were published at least as early as 1982 \cite{korill82}. It was then shown, measuring the kinematics of bulges and ellipticals, that some of these might be different stellar systems, at least from a dynamical viewpoint. These authors produced the seminal $V/\sigma$ vs. $\epsilon$ plot, where V stands for rotation velocity, $\sigma$ for velocity dispersion, and $\epsilon$ for ellipticty. In this diagram, the ellipticals and bulges used in this study occupy distinct loci. Further, theory indicates that those ellipticals occupy the locus where one finds stellar systems supported by anisotropic velocities, whereas those bulges are in the locus corresponding to isotropic oblate rotators \cite{bintre87}.

About ten years later, bulges showing a peculiar box/peanut morphology were added to this diagram \cite{kor93}. It was found that such bulges display a strong rotational support, which led to the conjecture that those bulges could actually be disks. Further work showed indeed that box/peanut bulges are the inner parts of bars that grow from the disk plane through dynamical instabilities\cite{kuimer95,burath05}.

More recent studies showed that there is yet a third kind of bulge apart from the classical ones and the box/peanut bulges. This is the disk-like bulge, which, as the name indicates, have properties similar to those of disks, and is thought to form with disk material via dynamical instabilities. Both the box/peanut and the disk-like bulge are called together as pseudo-bulges, but it is fundamental to distinguish between them \cite{ath05}. Although both, as opposed to the classical type, show important rotational support, they also have different properties (e.g. concerning their stellar population content) and after all a box/peanut bulge is just a part of a bar, which is itself a different structural component. Disk-like bulges are outliers in the Kormendy relation \cite{kor77} followed by classical bulges and ellipticals \cite{gad09}, and have younger stellar populations.

It is important to point out that a single galaxy can have all three kinds of bulges. It is not hard to imagine a classical bulge being formed, say, through mergers in a disk, which becomes dynamically unstable, forming a bar, which itself becomes unstable and originates a box/peanut bulge. Furthermore, gas and stars in the disk are brought to the galaxy center through the bar, thus producing a disk-like bulge. It is worth stressing that these three different stellar systems result in an excess of light above the inward extrapolation of the outer disk exponential surface brightness radial profile. Hence the notion that the three systems are photometric bulges.

The development of integral field spectrographs allowed a quantum leap on the study of bulge dynamics. The {\sc sauron} group \cite{ems+04,fal+06,gan+06} obtained 2D kinematic measurements of a number of bulges and ellipticals. These studies corroborated that bulges do rotate, and that some, presumably box/peanut bulges (i.e. bars), do that cylindrically, as expected for bars. They have also showed that the central regions of galaxies can be far more complex than one would naively expect. Examples of that are kinematically decoupled cores and velocity dispersion drops. Such complex displays can be the result of mergers, accretion or the presence of substructures such as inner disks.

Another structural component most relevant to understand the dynamics of disk galaxies is the thick disk, which shows different properties as compared to the thin disk. There is evidence that in massive galaxies thick disks rotate as thin disks, but in low mass galaxies, thick disks lag behind \cite{yoadal08}.

\section{Kinematical Measurements}

To shed light on bulge dynamics, it's important to try and get kinematical measurements from bulge stars with as little as possible contamination from other galactic structural components, in particular the disk. With this aim, we selected 14 galaxies with a range in bulge shape and luminosity, and with inclination angles very close to 90 degrees, i.e. edge-on galaxies, as strictly as practically feasible. We obtained long-slit spectra with the slit positioned along the disk plane and out of the plane of the disk, at a height which ensures us that contamination from the thin disk is low. However, an existing thick disk could still be present in these measurements. Another criterium to select the second positioning of the slit was to avoid visible dust structure. The observations were performed with the 2.3 m University of Arizona Steward Observatory Bok Telescope on Kitt Peak.

\begin{figure}
\includegraphics[keepaspectratio=true,width=5cm,clip=true]{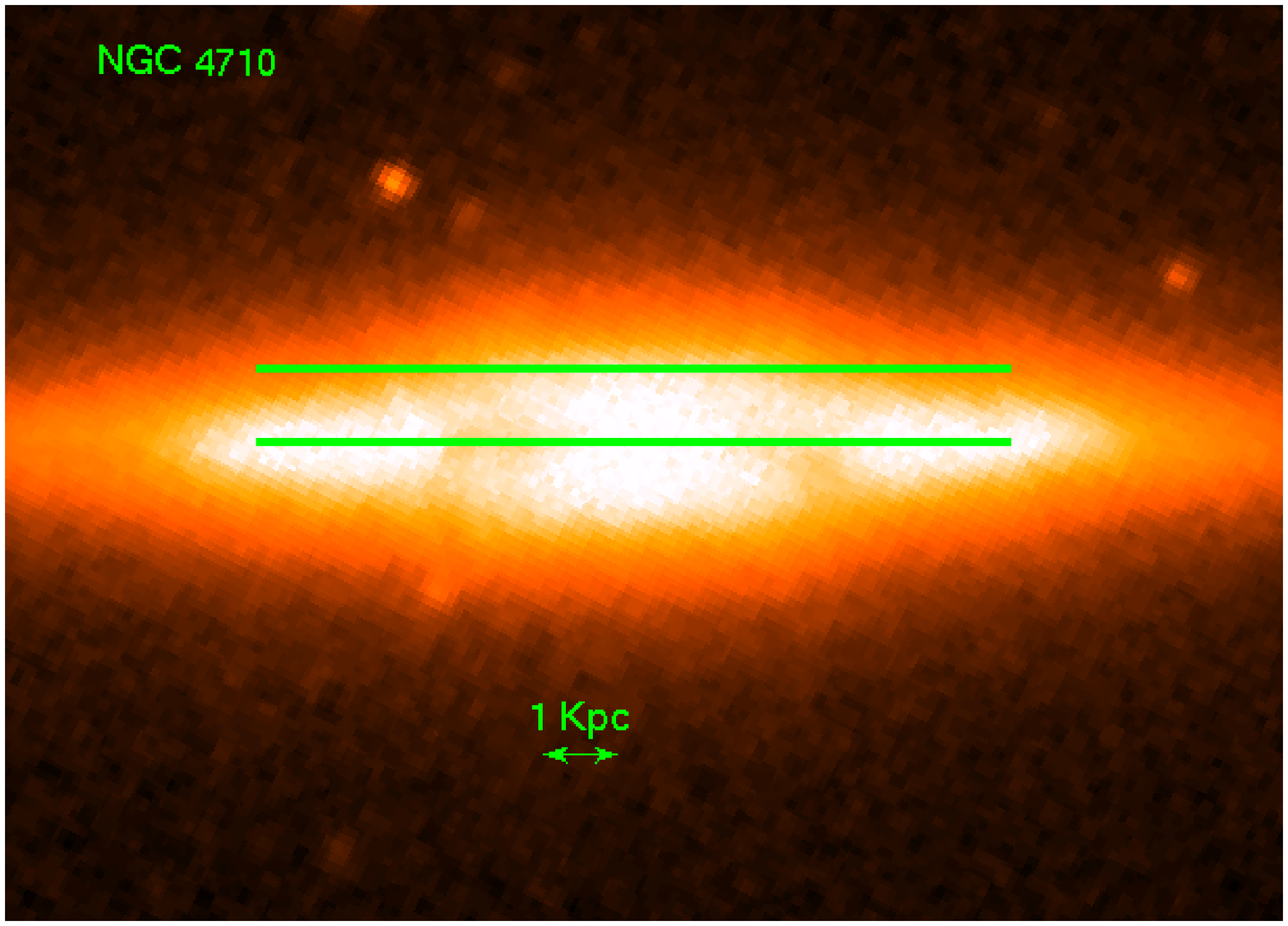}
\includegraphics[keepaspectratio=true,width=10cm,clip=true]{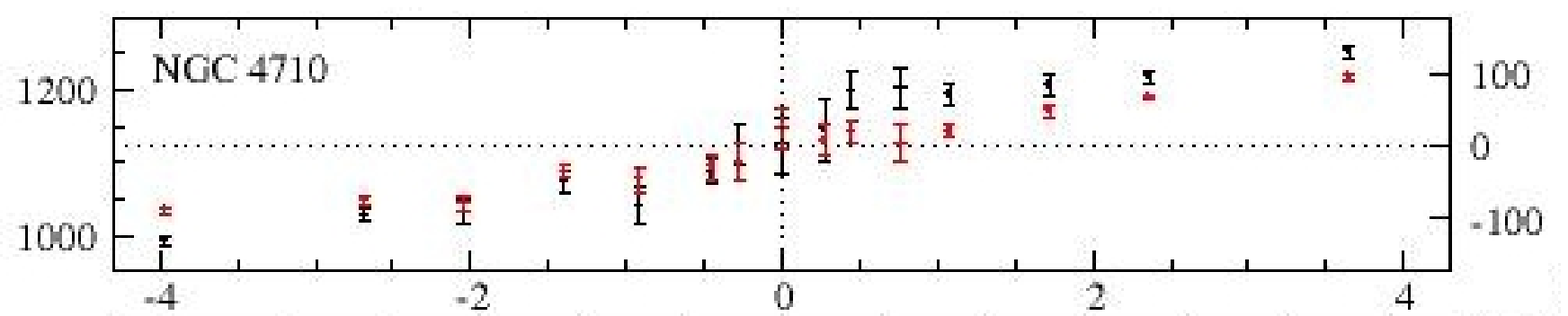}
\caption{Left: positioning of the long slits in the case of NGC 4710, in scale. Right: rotation curves obtained along the disk plane (in black) and above it (in red). Left y-axis is radial velocity in km/s, whereas right y-axis is radial velocity with systemic velocity subtracted. x-axis is radius in kpc.}
\end{figure}

All spectra were reduced and extracted through standard procedures, using tasks in the {\sc onedspec} and {\sc twodspec.longslit} packages in {\sc iraf}. We were particularly careful with flat-fielding, performing a response correction of the dome flat-fields to eliminate the continuum from the dome diffuse light, and a slit illumination
correction with twilight sky flat-fields. Fifteen spectra were extracted per slit position, along the slit, using the task {\sc kpnoslit.apall}.
For galaxies with a diameter at 25 $B$-mag. per sq. arcsec. smaller than $\approx$ 4 arcminutes, the
contribution from the sky was determined with the light in the outskirts
of the slit, where the light contribution
from the galaxy is much smaller, and subtracted from the data.
For larger galaxies, dedicated sky exposures were used. After calibration in wavelength, the uncertainty in velocity was estimated to be around $10-20$ Km s$^{-1}$. We found that the S/N per pixel in the spectra varies from around $10-20$ at about 1 arcminute from the center to $40-50$ at the center.

To obtain the radial velocity and velocity dispersion from each galaxy spectrum we
used the code written, tested and applied by us before \cite{gaddes05}. It performs a line profile
fitting in pixel space, assuming that the stellar velocity distribution is Gaussian, and uses
up to 5 different template stars. Many template stars were observed with the same instrumental set-up, and their spectra were reduced using the same procedure as for the galaxies' spectra. This approach minimizes errors from template mismatch. The spectral region used to measure the kinematical parameters
ranges from, approximately, the Mg {\sc i} triplet at $\lambda\approx 5175$ \AA\,
to the Na {\sc i} feature at $\lambda\approx 5893$ \AA. This region also includes relevant
lines in this respect, such as the Fe {\sc i} $+$ Ca {\sc i} lines at
$\lambda\approx 5265$ \AA\, and the Fe lines at $\lambda\approx 5328$ \AA. The H$\alpha$ and
[H$\beta$] lines, at 6563 and 4861 \AA, respectively, were excluded from the analysis to avoid
spurious results from the superposition of gaseous emission lines. Eventual emission lines,
such as [O {\sc iii}] (at $\approx$ 5007 \AA) were automatically excluded from the analysis by our
code, which ignores lines that are too discrepant in the galaxy
and template spectra.
Figure 1 shows our kinematical measurements for NGC 4710.

\section{Structural Analysis}

Light penetrating through the slits comes from stars at different structural components in the galaxy. At the slit in the disk plane, light comes predominantly from stars in the thin disk but, if present in the galaxy, also from stars in the bulge, bar and thick disk, each of these components influencing the spectrum differently at different parts along the slit. At the slit out of the disk plane, in the case of our observations, light from stars in the thin disk is significantly reduced, and it becomes dominated by other present structural components. To understand properly the kinematical measurements, i.e. to assess to which component these measurements correspond to, it is crucial to perform a detailed structural analysis. Only thus one is in a position to know which component (or components) dominate a given spectrum, and then properly interpret the rotation curves and velocity dispersion radial profiles obtained.

We thus used {\sc budda} \cite{desgaddos04,gad08} to decompose images of the galaxies in our sample into the different structural components present in each galaxy. Most of the images come from the Sloan Digital Sky Survey (SDSS) and we chose to use $i$-band images to avoid dust effects. For galaxies to which SDSS images are not available we used archival data retrieved through the NASA Extragalactic Database (NED), in optical red, near-infrared or mid-infrared passbands.

\begin{figure}
\includegraphics[keepaspectratio=true,width=7.5cm,clip=true]{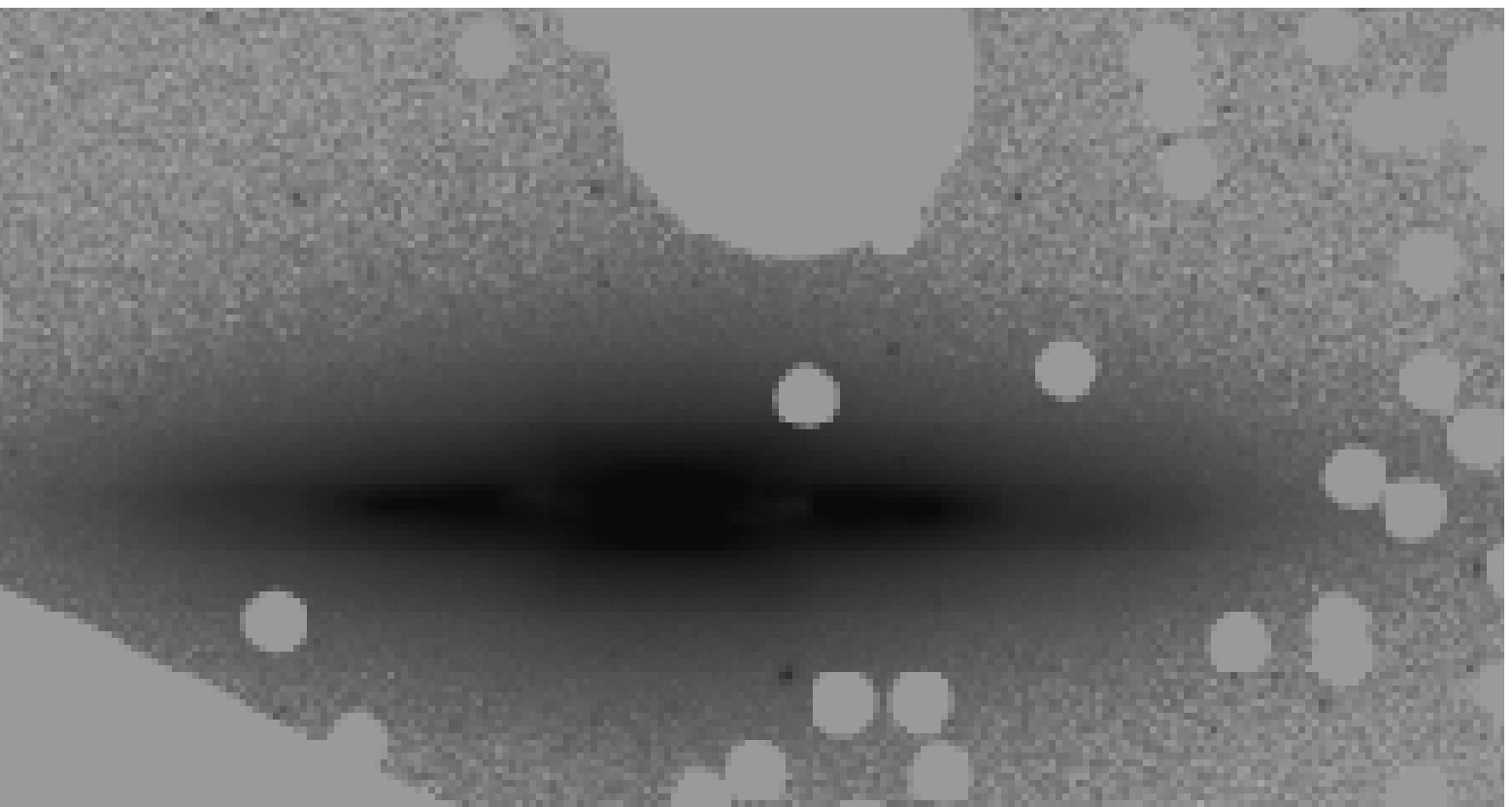}
\includegraphics[keepaspectratio=true,width=7.5cm,clip=true]{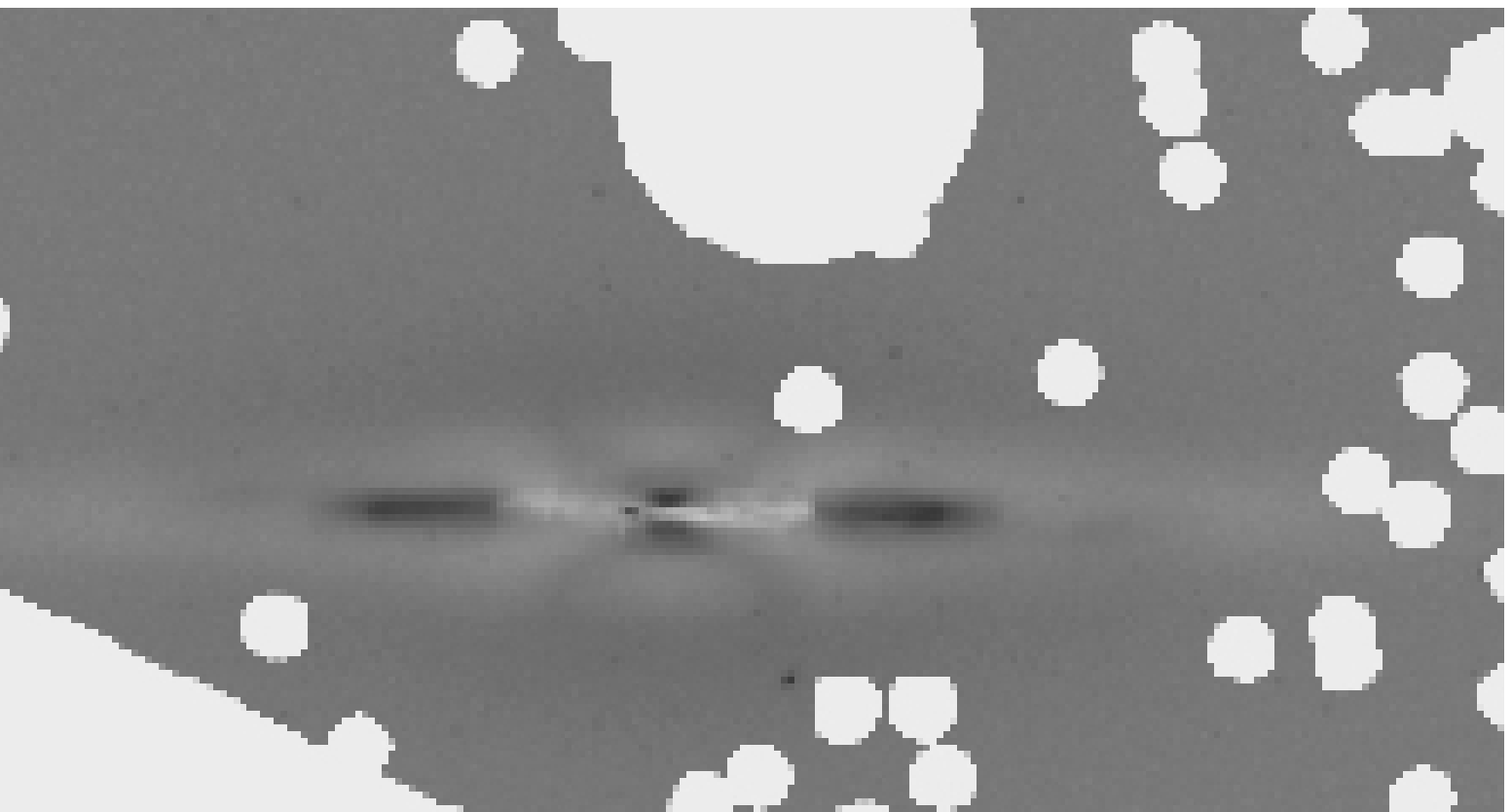}
\caption{Left: SDSS $i$-band image of NGC 4710 on a logarithmic scale. Right: residual image after subtraction of the {\sc budda} model including an edge-on disk, a bar and a very tiny exponential bulge (bulge-to-total ratio $B/T=0.001$). The X-shape in the residual image clearly indicates the presence of the peanut photometric bulge, which is just a structure in the inner part of the bar. Interestingly, NGC 4710 is classified in NED as a possibly ringed, unbarred S0 galaxy.}
\end{figure}

The {\sc budda} fits include an edge-on disk following Eq. (7) in \cite{vankru81}, and, when present, a S\'ersic bulge and a S\'ersic bar. These fits disclose many interesting features in the galaxies in our sample, as bars (revealed by an X-shape in the residual image), thick disks, and possibly, rings and spiral arms. Figure 2 shows the results from the decomposition of NGC 4710. The best model includes an exponential bulge with bulge-to-total ratio $B/T=0.001$, which suggests this can in fact be a bulgeless galaxy. The structure coming out of the disk plane is a bar, which is included in the model. The X-shape in the residual image strongly corroborates the presence of a bar. Also seen in the residual image are two thin structures at each side of the galaxy center, which are likely to be a ring (pseudo? broken?), or ansae at the bar ends. Interestingly, NGC 4710 is classified in NED as a possibly ringed, unbarred S0 galaxy.

Thus, in the case of NGC 4710, the structural analysis uncover that we cannot measure the rotation of the tiny exponential bulge (if it indeed exists), but only of the disk and the bar. In most other cases, we estimated the bulge rotation velocity from measurements within one bulge effective radius, where the bulge light dominates over other present components. For a few galaxies, as in the case of NGC 4710, we did not make such estimation, as we deemed that either the bulge was too small to properly measure its rotation with the current data, or the spectra are too much contaminated by light from a bar and/or a thick disk.

\section{Results and Discussion}

Table 1 shows the bulge S\'ersic index $n$, bulge rotation velocity $V$, estimated as described above, and $V/\sigma$, where $\sigma$ is the velocity dispersion at the galaxy center, for 8 galaxies. The table is sorted in a way that $n$ grows from left to right. For 3 galaxies, the measured rotation velocities within the bulge effective radius are consistent within the errors bars with no rotation whatsoever.

\begin{table}[h]
\caption{Bulge S\'ersic index $n$, bulge rotation velocity $V$ and $V/\sigma$ for 8 galaxies in our sample. $n$ grows from left to right. Clearly, exponential bulges have a more significant rotational support. In fact, some bulges show no significant rotation.}
\centering
\begin{tabular}{c|cccccccc}
\hline
\hline
$n$ & 1.2 & 1.4 & 1.8 & 2.0 & 2.1 & 2.3 & 2.6 & 4.3\\
$V$ (km/s) & 111 & 52 & $\approx$0 & $\approx$0 & $\approx$0 & 15 & 18 & 33\\
$V/\sigma$ & 0.57 & 0.58 & $\approx$0 & $\approx$0 & $\approx$0 & 0.12 & 0.17 & 0.21 
\end{tabular}
\end{table}

Table 1 clearly indicates that exponential bulges have significant rotational support, and are therefore disk-like bulges. Both the low S\'ersic index and the high $V/\sigma$ consistently lead to this conclusion. In contrast, bulges with higher S\'ersic indexes show little or no rotational support, and are thus classical bulges. Again, $n$ and $V/\sigma$ are consistent with this conclusion. Surprisingly, however, even bulges with $n$ as low as $\approx$ 2 are classical bulges, given their low $V/\sigma$. This reinforces the case for a very careful use of the bulge S\'ersic index as a discriminator between classical and disk-like bulges (see e.g. discussion in \cite{gad09}).

Our results show that classical bulges are dynamically similar to elliptical galaxies, with little rotational support, consistent with the idea that violent processes are to a large extent involved in the formation of both kinds of stellar systems. Furthermore, disk-like bulges show properties consistent with the picture in which these structures are formed from disk material (gas and stars), when disks become dynamically unstable and form non-axisymmetrical structures such as bars. Therefore, classical bulges and disk-like bulges are different stellar systems, with different physical properties and formation histories.

\section*{Acknowledgments}

I thank the organizers for their invitation and their warm hospitality. I am grateful to my collaborators in this study, R. de Souza, S. dos Anjos, R. Kennicutt and R. de Jong, for their invaluable work and input. This work has made use of SDSS, NED, 2MASS and ADS.

\end{document}